\documentclass[conference]{IEEEtran}
\usepackage{cite}
\usepackage{fancyhdr}
\usepackage{kantlipsum}
\fancypagestyle{plain}{
\fancyhf{}
\fancyhead[C]{Conference on \LaTeX} 

}
\usepackage{eso-pic} 
\usepackage{subfigure}
\usepackage{color}
\usepackage{amsmath,amssymb,amsfonts}
\usepackage{algorithmic}
\usepackage{graphicx}
\usepackage{multirow}
\usepackage[para,online,flushleft]{threeparttable}
\usepackage{textcomp}
\def\BibTeX{{\rm B\kern-.05em{\sc i\kern-.025em b}\kern-.08em
    T\kern-.1667em\lower.7ex\hbox{E}\kern-.125emX}}

\newcounter{RZNumberOfComments}
\stepcounter{RZNumberOfComments}

\begin{document}
\AddToShipoutPictureBG*{
\AtPageUpperLeft{
\setlength\unitlength{1in}
\hspace*{\dimexpr0.5\paperwidth\relax}
\makebox(0,-0.75)[c]{\textbf{2018 IEEE/ACM International Conference on Advances in Social Networks Analysis and Mining (ASONAM)}}}}

\title{Inspecting Interactions: Online News Media Synergies in Social Media\\
}

\author{\IEEEauthorblockN{Praboda Rajapaksha\IEEEauthorrefmark{1}\IEEEauthorrefmark{2}, Reza Farahbakhsh\IEEEauthorrefmark{1}\IEEEauthorrefmark{3}, No\"{e}l Crespi\IEEEauthorrefmark{1}, Bruno Defude\IEEEauthorrefmark{1}}
\IEEEauthorblockA{\IEEEauthorrefmark{1}\textit{Institut Mines-T\'el\'ecom, T\'el\'ecom SudParis, CNRS Lab UMR5157} \textit{Evry, France}\\.
\IEEEauthorblockA{\IEEEauthorrefmark{2}\textit{Uva Wellassa University, Passara Road, 90000, Badulla, Sri Lanka}}
\IEEEauthorblockA{\IEEEauthorrefmark{3}\textit{TOTAL SA., Strategy \& Innovation, EP/SG/ISB/STI, Data Analytics Competence Center (DACC)}}
\{praboda.rajapaksha, reza.farahbakhsh, noel.crespi, bruno.defude\}@telecom-sudparis.eu}
}

\maketitle

\label{sec:abstract}
\IEEEoverridecommandlockouts
\IEEEpubid{\parbox{\columnwidth}{\vspace{8pt}
\makebox[\columnwidth][t]{IEEE/ACM ASONAM 2018, August 28-31, 2018, Barcelona, Spain}
\makebox[\columnwidth][t]{978-1-5386-6051-5/18/\$31.00~\copyright\space2018 IEEE} \hfill} \hspace{\columnsep}\makebox[\columnwidth]{}}
\IEEEpubidadjcol 
\begin{abstract}
The rising popularity of social media has radically changed the way news content is propagated, including interactive attempts with new dimensions. To date, traditional news media 
such as newspapers, television and radio have already adapted their activities to the online news media by utilizing social media, blogs, websites etc. 
This paper provides some insight into the social media presence of worldwide popular news media outlets. 
Despite the fact that these large news media propagate content via social media environments to a large extent and very little is known about the news item producers, providers and consumers in the news media community in social  media.
To better understand these interactions, this work aims to analyze news items in two large social media, Twitter and Facebook.
Towards that end, we collected all published posts on Twitter and Facebook from 48 news media to perform descriptive and predictive analyses using the dataset of 152K tweets and 80K Facebook posts. 
We explored a set of news media that originate content by themselves in social media, those who distribute their news items to other news media and those who consume news content from other news media and/or share replicas. We propose a predictive model to increase news media popularity among readers based on the number of posts, number of followers and number of interactions performed within the news media community. The results manifested that, news media should disperse their own content and they should publish first in social media in order to become a popular news media and receive more attractions to their news items from news readers.

\end{abstract}


\section{Introduction}
News media have had a growing engagement in social media platforms for their news propagation. The term news media refers to the news sources that focus on delivering news to the public.
Based on the evidence available, there is a significant growth in the use of social media by news media to disseminate news \cite{97}, especially on Facebook and Twitter \cite{105} \cite{79}, where Twitter is more effective than Facebook in terms of audience reach \cite{97}. Interestingly, media news and news websites are the most common type of content shared in social media, more than other categories such as political, commentary, sports, movies\cite{33}.
In this perspective, one way for news media in this era to get ahead in the news media community is to spread news by publishing first and propagating accurate news on social media, thereby attracting more followers.
Hence, it is worthwhile to analyze the texts published by news media on Facebook and Twitter to identify how news stories are propagating among popular news media and to determine their interactions. It is yet not well-known how news media are interacting with each other in social media, and thus the results of this paper can be helpful to social media marketing managers in news organizations. 
Our motivation is to investigate news media content dissemination in social media in terms of news self-originators, news providers, and news consumers. These results provide insight into the behavior and the role of news aggregators and news agencies (those that collect news and sell to news media, including Agency France Presse-AFP, Associated Press-AP and Reuters) in social media.
 
Detecting news originator is key to investigating news media interactions. However, content originator detection is a challenging problem, especially in OSNs. We recently presented a framework to identify the content originators of textual content in OSNs, \textsl{ConOrigina} \cite{107}. ConOrigina detects a content creator by leveraging observable information such as their linguistic features and temporal behaviors (circadian typology). 
We explored the view that the consideration of the temporal changes of user behaviors in social media is significantly improved when it is used to identify content ownership. 
In a similar manner, we can apply the ConOrigina framework to evaluate the behaviors of news media in Twitter and Facebook based on thier published news items.

To that end, we used data mining, text mining and statistical methods to perform descriptive and predictive analyses on the dataset crawled from popular 48 news media. 
We defined 3 main metrics to quantify news media interactions and engagements in social media: 
1) \textsl{News self-originators}: those who originate content by themselves or publish fresh content, 2) \textsl{News providers}: those who distribute their content to other news media and 3) \textsl{News consumers}: those who mostly share replicated content that have been already published by other news media. 
One of the main outcomes of this work is to classify the sets of news media who act as news self-originators, news providers and news consumers based on their published content. 
Next, we analyzed which of the above metrics are correlated with the content popularity of the news media among news readers. 
Finally, we developed prediction models to predict news readers' reactions in the future using their historical behaviors such as the number of published posts, number of followers and total number of interactions performed with other news media. 
We determined from these prediction models that, in order to increase reader reactions or popularity among news readers, news media must publish fresh content, they have to deliver content to other news media to increase their interactions and they also have to increase the number of followers on their social media pages.


\section{Dataset and Framework Description}
\label{sec:dataset_description}

This section begins by describing the dataset used in this study for identifying the behaviors of news media in terms of their published content. 
There was no previous work that has analyzed news media popularity based on content originality and on their interactions. To this end, we have identified 48 news media popular worldwide and their user accounts in Facebook and Twitter. The framework described in section \ref{ConOrigina} was then used to detect news originators.

\subsection{Data Collection}
The dataset used in this study is based on the top 48 most popular news media sites (English edition) ranked by Alexa\footnote{http://www.alexa.com/topsites/category/Top/News}, 
retrieved on 2\textsuperscript{nd} May 2017. 
The list is based on the amount of traffic, the number of unique visitors, recorded for each news media. 
We manually identified a set of active and authorized news media user accounts that officially represent the news media sites in both Facebook and Twitter. 
We considered a single account per news media: the most popular news media account with the highest number of followers compared to other accounts who represent the same news media. 

Next, for each news media user, we executed Facebook and Twitter crawlers (for one month: 8\textsuperscript{th} May - 8\textsuperscript{th} June 2017) to extract timeline posts, respective timestamps, and number of likes, shares, comments, retweets and favorites. 

\begin{table}[]
\centering
\caption{Dataset description}
\begin{tabular}{l|c|c|c|}
\cline{2-4}
                                  & \multicolumn{1}{l|}{\textbf{Total \#user-IDs}} & \multicolumn{1}{l|}{\textbf{\#followers(avg)}} & \multicolumn{1}{l|}{\textbf{Total \#posts}} \\ \hline
\multicolumn{1}{|l|}{\textbf{Twitter}} & 48                                             & 2836K                                          & 152K                                        \\ \hline
\multicolumn{1}{|l|}{\textbf{Facebook}} & 48                                             & 3630K                                          & 80K                                         \\ \hline
\end{tabular}
	\vspace{-0.4cm}

\label{table:dataset_brief_description}
\end{table}


\subsection{Dataset Description}
A brief description of the dataset acquired from Twitter and Facebook is shown in Table \ref{table:dataset_brief_description}, contains information about the average number of followers and total number of posts shared. It is clear that the largest number of followers are from the news media on Facebook compared with them on Twitter. 
In Facebook, highest number of posts (10.6K) are shared by the \textsl{Indianexpress} and \textsl{Hindustantimes(6.7K)}. 
In Twitter, top 10 news media that have published large number of news items are \textsl{Bloomberg, Cnn, Indiatimes, Foxnews, Hindustantimes, Indianexpress, Theguardian, Thehill, Time}, and\textsl{ Economictimes}.
The least number of posts were shared by \textsl{Reddit} on Facebook.

\begin{figure}
\centering
\footnotesize
\subfigure[\#posts-FB \& TW.]{%
\label{fig:singleUserdistro}%
\includegraphics[width=1.07in]{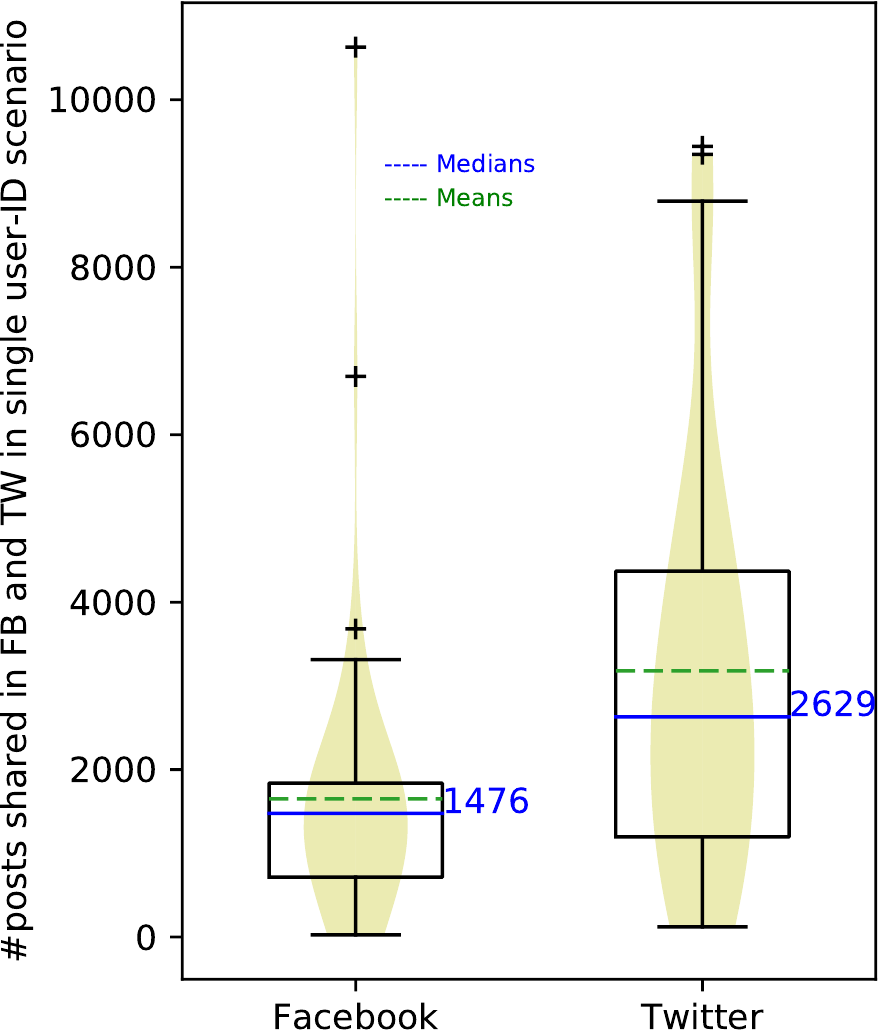}}%
\qquad
\subfigure[Reactions-TW.]{%
\label{fig:boxplot_consumerreactionsTwitter}%
\includegraphics[width=0.98in]{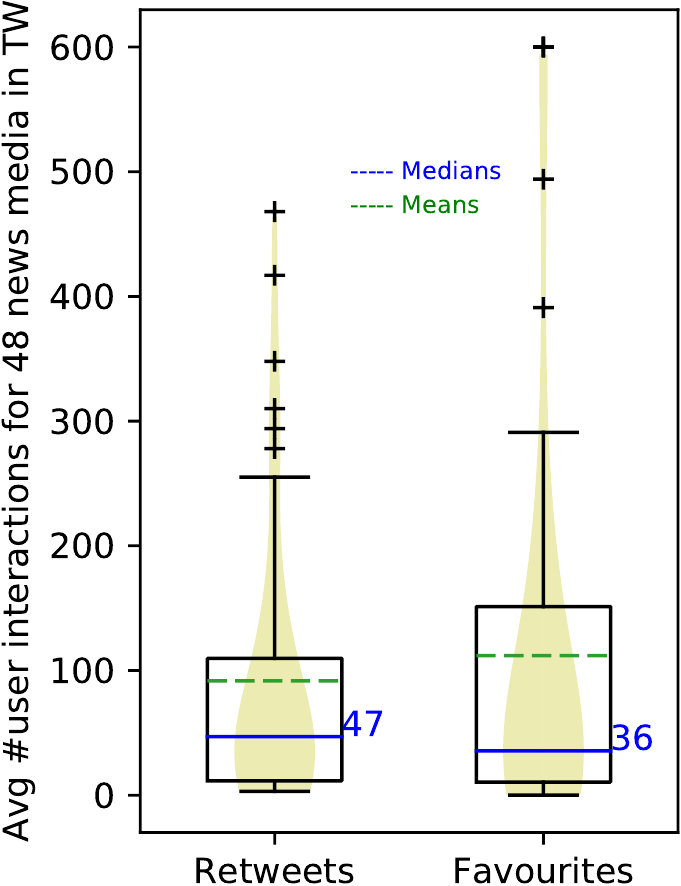}}%
\qquad
\subfigure[Reactions-FB.]{%
\label{fig:boxplot_consumerreactionsFacebook}%
\includegraphics[width=0.98in]{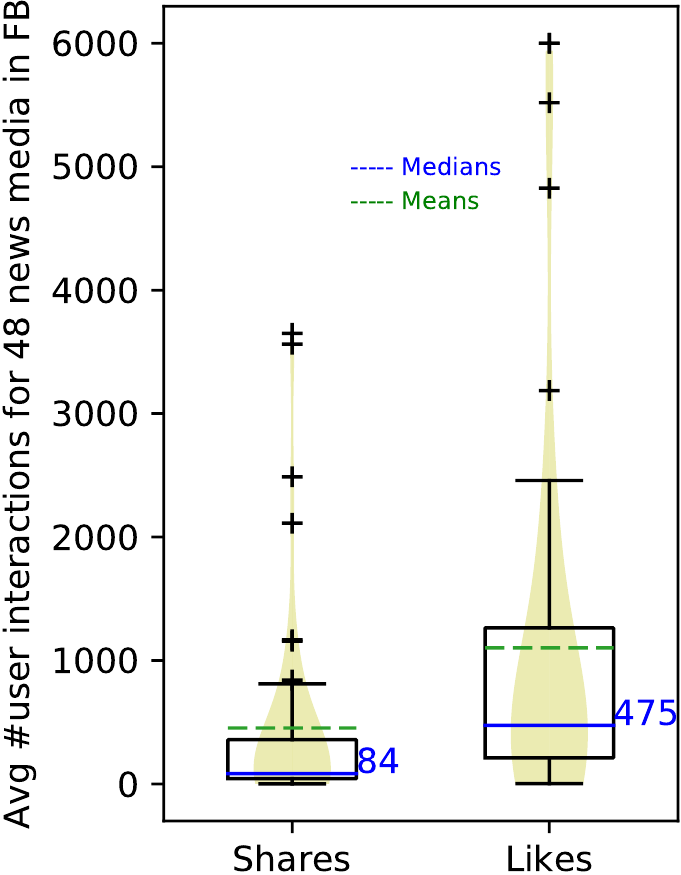}}

\vspace{-0.25cm}
\caption{The distributions of the number of posts shared by the news agencies in social media and their news reader interactions.}
\label{fig:Alldistro}
\vspace{-0.2cm}
\end{figure}

The violon plot in Figure \ref{fig:Alldistro} shows further interpretation on the dataset. Figure \ref{fig:singleUserdistro} illucidates the distribution among number of news items posted on Facebook and Twitter by all news media, which illustrates the abstract representation of the probability distribution of the dataset based on the symmetrical kernel density estimation (KDE). Figure \ref{fig:singleUserdistro} clearly presents that in average, number of tweets shared by these news media is higher than that on Facebook. 
In spite of the fact that the highest number of followers are from Facebook news media accounts compared to the same set of the news media representatives in Twitter (as indicated in Table \ref{table:dataset_brief_description}), number of posts shared by them on Facebook is lower than the number of tweets. Some preliminary work carried out in the recent years have also proven the same that Twitter has been widely used as a source of news than Facebook \cite{97},\cite{98}.

The distribution of the average number of reader reactions for all the posts in our dataset is shown in Figure \ref{fig:Alldistro} with the box-plots for Twitter (Figure \ref{fig:boxplot_consumerreactionsTwitter}) and Facebook (Figure \ref{fig:boxplot_consumerreactionsFacebook}). 
It is apparent from Figure \ref{fig:boxplot_consumerreactionsFacebook} that news consumers reacted more on Facebook posts than on tweets and it is obvious that the largest number of followers are from Facebook (Table \ref{table:dataset_brief_description}). 

As a summary, based on our analysis, 77\% of the news media were more active on Twitter than on Facebook, as manifested in the previous works \cite{97} \cite{98}. \textsl{Indianexpress} and \textsl{Hindustantimes} were used both social networks very actively. A few others, \textsl{Reddit and Ipsnews}, have only a minor engagement with social media compared to other news media, and the remainder mostly use either Facebook or Twitter exclusively. 
Apart from that, although news media sites in Facebook do not share as much content as in Twitter, number of reader reactions in Facebook is considerably higher than that of Twitter.

\subsection{Framework Description}
\label{ConOrigina}

The analysis of the news media dataset enable us to investigate interactions among them on social media, in particular, which news media produce, distribute, and consume the news items. The detection of the content originator of texts is an important asset to observe information propagation patterns among different news media especially in the context of sharing the replicas by different news media. In that regard, the framework ConOrigina\footnote{The framework will be public to the community for further research in
GitHub in the camera ready version. Contact authors for more information.} \cite{107} is capable of detecting content originator of news items. The ConOrigina framework was designed to identify textual content originator in OSNs based on the SCAP method \cite{136} to detect linguistic patterns and online circadian behavior \cite{63} to detect temporal patterns of the user. 
The ConOrigina uses four distinct phases; i) Data-cralwer, ii) Pre-processor, iii) Feature-extractor, and iv) Author-analyzer. 

\textit{Data-crawler} extracts relevant information from OSNs to build a knowledge base, mainly author details, and content attributes. The use of raw data in ConOrigina reduces the accuracy of the results, as we crawled data from different social media that use different content properties (e.g, length) and user profiles. 
Therefore, \textit{pre-processor} in ConOrigina consider two criteria to filter raw dataset; 
1) remove posts with word-count=0 and word-count$<$3 and
2) remove URLs. 
\textit{Feature-extractor} receives as input a dataset classified in the pre-processor phase and outputs an average similarity index per user based on their similar writing patterns by applying SCAP method to detect pairwise similarity between two different texts/posts. \textit{Author-analyzer} phase uses Jenks optimization algorithm and TFV method \cite{107} (uses online circadian patterns) to classify potential authors for a given text. 

\section{News Media Interaction Analysis}
\label{sec:news_agency_analysis}

It is worthwhile to better understand news dissemination patterns among news media in Facebook and Twitter in terms of their interactions. Interactions can be inspected based on the shared self-originated content and replicated content as explained in the following sections.

\subsection{Detection of the Originated and Replicated News Items}
In order to analyze news media interactions in social media, first we need to identify the originated and replicated content they shared. The ConOrigina\cite{107} framework was first executed on the pre-processed dataset, as detailed in section \ref{sec:dataset_description}.
We amalgamated pre-processed tweets and Facebook posts together, in a total of 232K posts, to detect the owner of each post using our ConOrigina. ConOrigina evaluates pair-wise syntactic similarity among each post and outputs a possible news media that originated a particular post. 
As a result, we can identify the potion of the original posts and replicas shared by all news media. The detection of the replicas helps to investigate news consumers and distributors in social media.

\begin{figure}[h]
	\centering
	\includegraphics[width=3.7in]{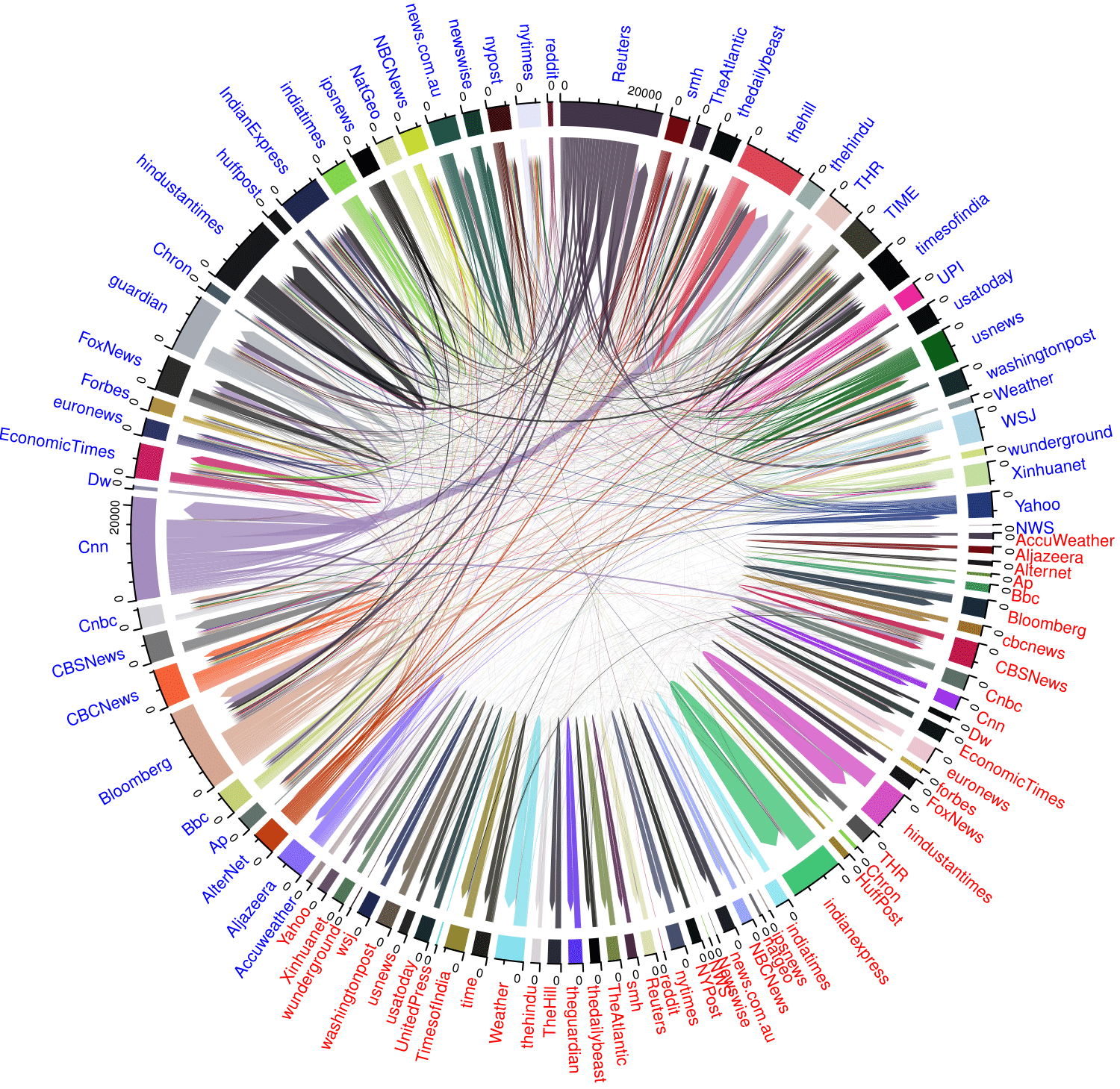}
				\vspace{-0.7cm}
		\caption{The news propagation patterns and user interactions among 48 different news agencies considered in our dataset. TW user-IDs are represented in blue and FB user-IDs are represented in red. The large arrows indicate the directional flow of the content. }
			\vspace{-0.4cm}

	\label{fig:News_prop_TW_FB}
\end{figure}

\subsection{News Agencies Interactions}

This section explains content interactions among news media based on their published posts. 

We measure the number of interactions among news media in terms of three different measures: \textit{SelfLink, DispersedLinks} and \textit{AcquiredLinks}. 
For instance, three measures of news media A can be expressed as: 
\textit{ i) SelfLink (A$\,\to\,$A)} - the portion of the self-originated content by A;
ii) \textit{DispersedLinks (A$\,\to\,$B)} - the portion of the content originated by A that is dispersed to B; and, 
iii) \textit{AcquiredLinks (B$\,\to\,$A)} - the portion of the content acquired by A that originated in B (reverse of DispersedLinks).
Using these parameters, we build a model as follows to convey the total number of interactions (`In') performed by each news media. W stands for the weight and thus $W_{SelfLink}$ represents the weight of the SelfLink.  

\textit{`In'} $= \sum  (W_{SelfLink}+W_{DispersedLinks}+W_{AcquiredLinks}) $ \\

The interaction patterns among news media users are shown in Figure \ref{fig:News_prop_TW_FB}. The sectors correspond to each news media. The width of a sector is proportional to the number of interactions that a particular news media performs with other users. The sector sizes in Figure \ref{fig:News_prop_TW_FB} 
clearly indicate the value of `In' for each news media. Grid colors represent sectors, and link colors, as with the grid colors, represent different new media. The position of the starting point of the link is shorter than the other end to show that the link is moving from one news media (DispersedLinks) and being received/consumed by another (AcquiredLinks). The large arrows in the graph indicate the direction of the information flow. The names of the news media are listed in alphabetical order for ease of comprehension: news media on Twitter are labeled in blue and those of Facebook are indicated in red. 

As depicted in Figure \ref{fig:News_prop_TW_FB}, in majority of the news media, the largest `In' appears with their Twitter user accounts. We can deduce from Figure \ref{fig:News_prop_TW_FB} that news media are more active in Twitter than in Facebook in terms of total interactions performed with others, and that a majority of them have shared self-originated content on Facebook and have a limited number of interactions with others in both Facebook and Twitter. 

Among all the users in our dataset, the highest `In' values, more than 18K, were performed by \textsl{Reuters} and \textsl{Cnn} in Twitter. The AcquiredLinks of these two news media compared with those of others are negligible.
For instance, the weight of the AcquiredLink of \textsl{Reuters} (Cnn$\,\to\,$Reuters) is 3.3\% from its `In' and, that of \textsl{Cnn} (Reuters$\,\to\,$Cnn) is only 0.21\%. Therefore, Cnn and Reuters are more active in Twitter since they show highest `In' values among all other news media. However in Facebook, the largest `In' value appeared from \textsl{IndianExpress}, about 6K. As a result, we can conclude that \textsl{IndianExpress} is the most active news media in Facebook.

\subsubsection{\textit{Strong self-content publisher}}
A strong self-content publisher has a considerable weight for its SelfLink compared with the weights of its DispersedLinks and AcquiredLinks. Strong self-content publishers are the fresh content producers.

There are 5 news media that have exhibited a large portion of the self-originated content on Twitter, more than 80\% of the content shared by them are self-originated, as shown in Figure \ref{fig:News_prop_TW_FB}. These are \textsl{Accuweather, Alternet, Cnn, Dw} and \textsl{Newswise}. In contrast, nearly all of the news media on Facebook shared posts $>$ 80\% as self-originated content. This observation made visible in Figure \ref{fig:News_prop_TW_FB}, shows that the majority of the news media shared their own content on Facebook. 

To conclude this section, the results show that, 5 news agencies: \textsl{Accuweather, Alternet, Cnn, Dw}, and \textsl{Newswise} have published more fresh content than the others, thus acting as real news items producers.

\subsubsection{\textit{Strong content provider}}
A strong content providers exhibit a large portion of the number of DispersedLinks than its portion of the SelfLink and DispersedLinks. They are the content leaders in the news media community in social media.

Based on our analysis, only 10 news media exhibit $>$ 70\% for the weight of DispersedLinks. These are \textsl{Accuweather, Alternet, Cnn, Dw, Ipsnews, Nationalgeographic, Reddit, Reuters, UPI} and \textsl{Weather.gov}. These news media can be considered as the content providers in Twitter in particular, \textsl{Ipsnews} and \textsl{Alternet} exhibit $>$ 95\% of DispersedLinks weight, and as indicated in Figure \ref{fig:News_prop_TW_FB} almost all of their links are outward links.
A notable relationship in Figure \ref{fig:News_prop_TW_FB} is that the largest DispersedLink is between Twitter representatives of Cnn and TheHill (cnn$\,\to\,$TheHill). This indicates that 42\% of the `In' values of TheHill are performed with Cnn, which further conveys that the majority of the posts published by TheHill are replicas that were initially shared by Cnn. In contrast, there is no any important relationship among news media present in Facebook in terms of their DispersedLinks content.

To sum up, we explored the top 10 content providers in the news media community (\textsl{Accuweather, Alternet, Cnn, Dw, Ipsnews, Nationalgeographic,Reddit, Reuters, UPI} and \textsl{Weather.gov}). The news shared by these news media are distributed to other news media. 
Most importantly, it is noted that the strong self-content publishers in the previous section are also strong content providers, except for \textsl{Newswise}.  
 
\subsubsection{\textit{Strong content consumer}}
The term strong content consumer is used to group news media that share many replicas (have a high number of AcquiredLinks). 

As shown in Figure \ref{fig:News_prop_TW_FB}, some news media users present a large number of AcquiredLinks with respect to their number of DispersedLinks. 
Based on the results, the top 10 content consumers in Twitter have $>$ 70\% for their weight of the AcquiredLinks: \textsl{Ap, Chron, Hollywoodreporter, Indianexpress, Timesofindia, Nytimes, Theatlantic, Thedailybeast, Usatoday}, and \textsl{Washingtonpost}. Among them \textsl{Ap, Usatoday} and \textsl{Indianexpress} have $>$ 82\% AcquiredLinks weights. 
The statistics show that \textsl{Ap} has the least number of DispersedLinks and also has insignificant levels of self-links, at 0\% and 5.6\% respectively. These findings add substantially to our conclusion that \textsl{Ap} is a strong content consumer, as it has a high weight for its AcquiredLinks. The precise representation of this discussion is also presented in Figure \ref{fig:News_prop_TW_FB} where \textsl{Ap} in Twitter is shown with all the links as incoming links.
Apart from that, even though \textsl{Usatoday} posted a considerable amount of tweets (2587), only 6.3\% were allocated for its SelfLink and DispersedLinks, with the remainder assigned for AcquiredLinks. More interestingly, the news media with the highest number of posts in our dataset(10.6K), \textsl{Indianexpress}, contains a SelfLink value of 11\%, while all its remaining links are AcquiredLinks with an insignificant number of DispersedLinks. This finding is also made visible in Figure \ref{fig:News_prop_TW_FB}. There are no significant results for strong content consumers in the news media on Facebook.

To summarize, we identified the top 10 strong content consumers in the news media community in social media as \textsl{Ap, Chron, Hollywoodreporter, Indianexpress, Timesofindia, Nytimes, Theatlantic, Thedailybeast, Usatoday}, and \textsl{Washingtonpost}. The content published by these news media claimed to have the highest number of replicas that were previously posted by the remaining news media in our dataset.

\subsection{News Media Popularity}
Marketers follow different strategies to measure the type of interactions from their customers in social media. The simplest metric is to track the number of likes and shares received and their audience growth/rate of followers. This section aims to analyze news media in terms of the reader reactions received on shared posts aiming to build a predictive model to increase future reader reactions.

\subsubsection{News reader reactions}

As illustrated in Figure \ref{fig:boxplot_consumerreactionsFacebook}, news readers on Facebook exhibit a larger portion of \#likes (avg-1455) than \#shares (avg-454), and similarly in Twitter, in Figure  \ref{fig:boxplot_consumerreactionsTwitter}, the favorite count (avg-133) is higher than the re-tweet count (avg-92). 
The top 10 popular news media in Twitter in terms of both \#favorites and \#re-tweets are \textsl{FoxNews, Cnn, Nationalgeographic, Nytimes, Washingtonpost, Thehill, Ap, Huffingtonpost, Bbc.co.uk}, and \textsl{Aljazeera}. Among them, the highest \#re-tweets count is presented with \textsl{FoxNews} and \textsl{Cnn}, which is about 1K and the largest \#favorites count, aproximatly 0.5K, is also exhibited in \textsl{FoxNews} and \textsl{Cnn}. 
Whereas in Facebook, top 10 news agencies with large number of likes and shares are \textsl{FoxNews, Bbc.co.uk, Cnn, Nationalgeographic, Aljazeera, Huffingtonpost, TheHill, Nytimes, Indiatimes}, and \textsl{Usatoday}. The \textsl{FoxNews} has aquired 12K \#likes and 3.7K \#shares count in Facebook, \textsl{Bbc.co.uk} has received the second largest counts of \#likes and \#shares, 9K and 3.6K respectively.

To conclude here, from the above discussion, we discovered top 10 news media in Twitter and Facebook with regard to their reader reactions received on the published news items. We identified 8 news media those that shared widely popular news items in both Facebook and Twitter: \textsl{FoxNews, Bbc.co.uk Cnn, Nationalgeographic, Nytimes, Thehill, Huffingtonpost}, and \textsl{Aljazeera}. It is noticeable that the \textsl{FoxNews} news items received much more attractions from news readers than other news agencies contents in both Facebook and Twitter.  


\subsubsection{Reader Reactions Predictive Model}
\label{sec:prediction_model}
This section focus on exploring an existence of a correlation between news media content popularity with other attributes such as number of posts and self-originated, acquired and dispersed contents.

\begin{figure}
	\centering
	\subfigure[Correlation on TW attributes.]{%
		\label{fig:correlation-TW}%
		\includegraphics[width=1.8in]{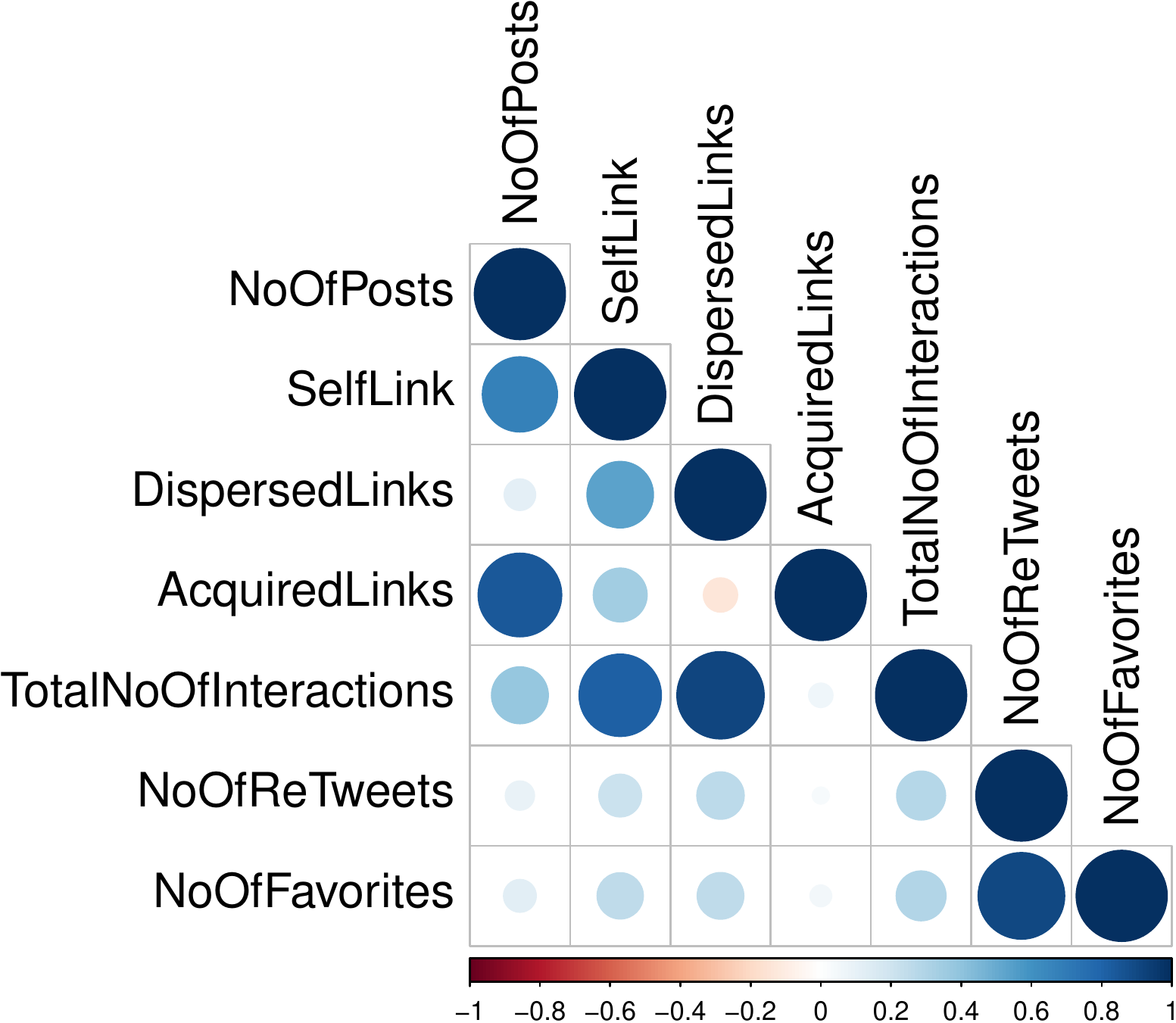}}%
	\subfigure[Correlation on FB attributes.]{%
		\label{fig:ccorrelation-FB}%
		\includegraphics[width=1.8in]{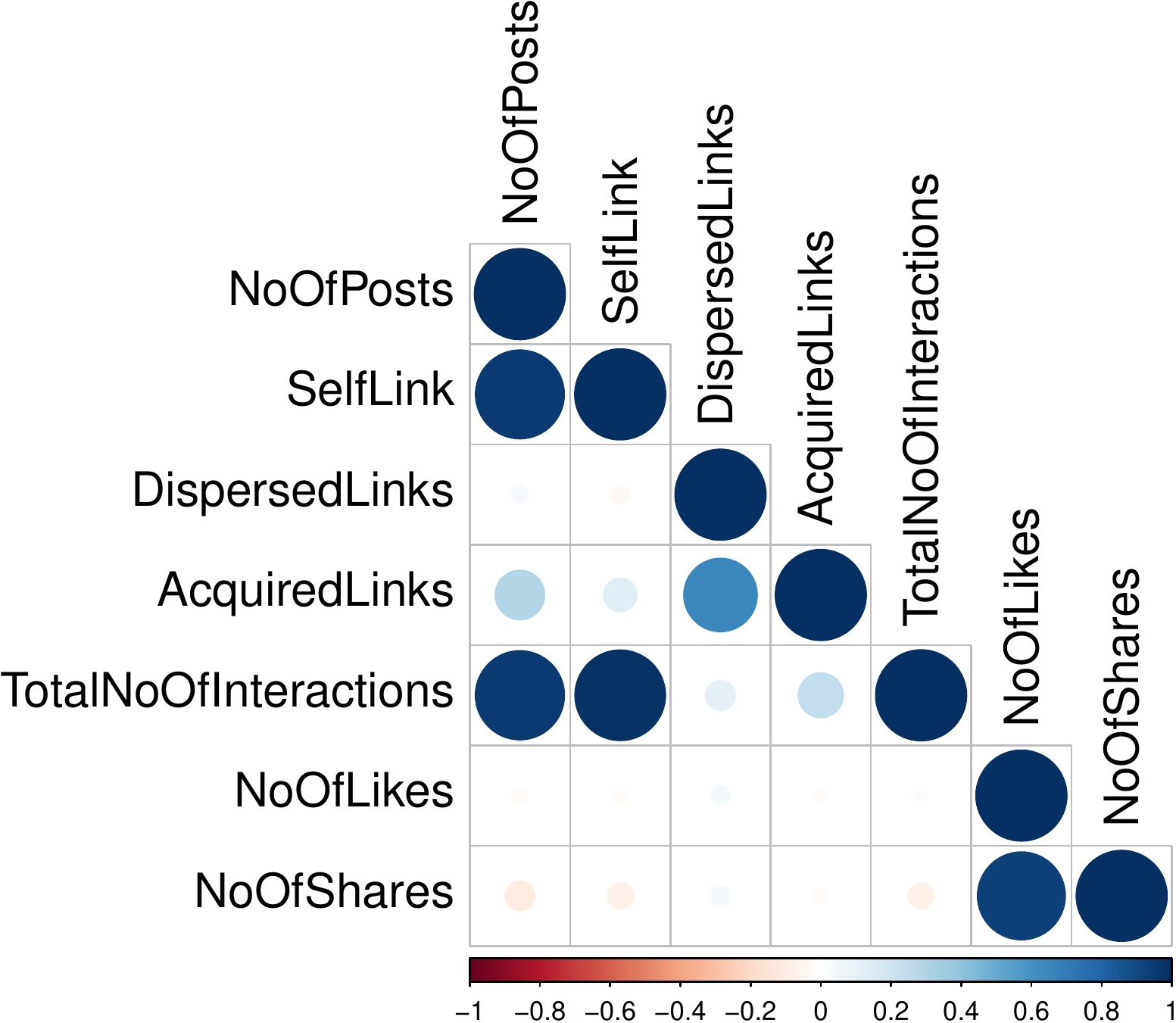}}%
	\vspace{-0.2cm}
	\caption{The correlation plots show the association between different attributes identified on the news agencies that are used in this study such as \#posts, SelfLink, DispersedLinks, AcquiredLinks and reader reactions.}
	\label{fig:correlation-graphs}
		\vspace{-0.4cm}

\end{figure}

Figure \ref{fig:correlation-graphs} exhibits correlation graphs (based on the Pearson correlation coefficient) for the variables of number of posts, reader reactions and weights of SelfLink, DispersedLinks, AcquiredLinks of each news media in our dataset for both Facebook (Figure \ref{fig:ccorrelation-FB}) and Twitter (Figure \ref{fig:correlation-TW}) seperatly.
In Facebook, \#news items posted by news media are highly correlated with their self-originated content and `In'. However in Twitter, significant positive correlation exists between `In' and $W_{SelfLink}$ and $W_{DispersedLinks}$. Also, $W_{SelfLink}$ and $W_{DispersedLinks}$ have strong positive correlation with the \#news. 
Therefore, 2 major attributes that are positively correlate with reader reactions are \#posts and `In'(propositional to $W_{SelfLink}$ + $W_{DispersedLinks}$).


We built 2 linear regression models, as illustrated in Table \ref{regression-model}, considering 3 major attributes: \#posts, ’In’ and \#followers, to predict future news reader reactions.
We can observe that among three response variables, only the \#followers and `In' have a p-value closer to 0, and they are in the range of p-values of 5\%. Consequently, we can see a relationship between \#followers and `In' with \#retweets and \#favorites, while \#posts do not have much effect on the predictor variables. 

These findings conclude that, in order to increase the content popularity of the news agencies among readers they should publish the content that they themselves originated and they should distribute those content to other news agencies acting as content providers. Further, higher \#followers in a news group could exhibits large \#reader-reactions.

\begin{table}[t]
	\centering
	\scriptsize 
	\caption{Regression models for re-tweets and favourites}
	\label{regression-model}
	\begin{threeparttable}
		\begin{tabular}{ll|l|l|l|l|}
			\cline{3-6} 
			&                         & \textbf{Estimate} & \textbf{Std. Error} & \textbf{t value} & \textbf{Pr(\textgreater t)} \\ \cline{2-6} \cline{1-1}
			\multicolumn{1}{|l|}{\multirow{5}{*}{\textbf{Model re-tweet}}} & \textbf{(Intercept)}    & 30.5          & 28.1            & 1.085            & 0.284                     \\ \cline{2-6} 
			\multicolumn{1}{|l|}{}                                         & \textbf{\#posts}        & -0.011         & 0.008            & -1.351           & 0.184                    \\ \cline{2-6} 
			\multicolumn{1}{|l|}{}                                         & \textbf{\#followers}    & 0.001          &0.001             & 2.821            & 0.007**                  \\ \cline{2-6} 
			\multicolumn{1}{|l|}{}                                         & \textbf{In} & 0.012          & 0.005             & 2.22             & 0.032*                   \\ \cline{2-6} 
			\multicolumn{1}{|l|}{}                                         & \multicolumn{5}{c|}{R-squared: : 0.221 and P-value: 0.011}                                                     \\ \cline{2-6} \cline{1-1}
			\multicolumn{1}{|l|}{\multirow{5}{*}{\textbf{Model favorite}}} & \textbf{(Intercept)}    & 7.72          & 59.100            & 0.131            & 0.897                     \\ \cline{2-6} 
			\multicolumn{1}{|l|}{}                                         & \textbf{\#posts}        & -0.019         & 0.017            & -1.067           & 0.292                     \\ \cline{2-6} 
			\multicolumn{1}{|l|}{}                                         & \textbf{\#followers}    & 0.000          & 0.000            & 2.373            & 0.022*                    \\ \cline{2-6} 
			\multicolumn{1}{|l|}{}                                         & \textbf{In} & 0.023          & 0.011            & 2.095            & 0.042*                     \\ \cline{2-6} 
			\multicolumn{1}{|l|}{}                                         & \multicolumn{5}{c|}{R-squared: 0.186 and P-value: 0.027}                                                       \\ \cline{2-6} \cline{1-1}
		\end{tabular}
		\begin{tablenotes}
			\item[] Significant codes:  0 `***' 0.001 `**' 0.01 `*' 0.05 `.' 0.1 ` ' 1
		\end{tablenotes}
		
	\end{threeparttable}
		\vspace{-0.4cm}

\end{table}

\section{Conclusion}
\label{sec:conclusion}

In this study, we analyzed 48 news media engagements in social media by crawling public information from their Facebook and Twitter accounts, 152K tweets and 80K Facebook posts.
We used our implemented framework ConOrigina to distinguish self-originated content from replicas to identify news origination, news dissemination, and news consumption interactions. 
We first explored the news media who published the largest \#posts. Next, with the results of ConOrigina, we discovered news item producers, consumers and providers (distribute to others). Almost all the news media who identified as self-content publishers (news producers) are also classified as content providers namely;\textsl{Accuweather, Alternet, Cnn, Dw, Ipsnews, Nationalgeographic, News.yahoo, Reddit, Reuters and Weather.gov}. Next, we determined the set of news media who mostly published replicas; include \textsl{Ap, Chron, Hollywoodreporter, Indianexpress, Timesofindia, Nytimes, Theatlantic, Thedailybeast, Usatoday and Washingtonpost}.

Apart from that, predictive model showed that the \#followers and total \#interactions (in terms of self-originated and dispersed content) are correlated with the reader reactions. Therefore, news media should disperse their own content and they should publish first in social media in order to become a popular news media and receive more attractions to their news items from news readers.

\bibliographystyle{IEEEtran}  
\bibliography{citation}



\end{document}